# Multimodal Plasmonics in Fused Colloidal Networks

Alexandre Teulle[1], Michel Bosman[2]*, Christian Girard[1], Kargal L. Gurunatha[1], Mei Li[3], Stephen Mann[3], Erik Dujardin[1]*

[1] CEMES CNRS UPR 8011, 29 rue J. Marvig, 31055 Toulouse Cedex 4, France.

[2] Institute of Materials Research and Engineering, A*STAR (Agency for Science, Technology and Research), 3 Research Link, 117602 Singapore.

[3] Centre for Organized Matter Chemistry, School of Chemistry, University of Bristol, Cantocks Close, Bristol BS81TS, United Kingdom.

* Correspondence to: erik.dujardin@cemes.fr, michel.bosman@gmail.com

**Abstract**

**Harnessing the optical properties of noble metals down to the nanometer-scale is a key step towards fast and low-dissipative information processing. At the 10-nm length scale, metal crystallinity and patterning as well as probing of surface plasmon (SP) properties must be controlled with a challenging high level of precision. Here, we demonstrate that ultimate lateral confinement and delocalization of SP modes are simultaneously achieved in extended self-assembled networks comprising linear chains of partially fused gold nanoparticles. The spectral and spatial distributions of the SP modes associated with the colloidal superstructures are evidenced by performing monochromated electron energy loss spectroscopy with a nanometer-sized electron probe. We prepare the metallic bead strings by electron beam-induced interparticle fusion of nanoparticle networks. The fused superstructures retain the native morphology and crystallinity but develop very low energy SP modes that are capable of supporting long range and spectrally tunable propagation in nanoscale waveguides.**

Surface plasmons (SPs), which are collective oscillations of conduction electrons, are regarded as a promising gateway to on-chip sub-wavelength electro-optical devices.[1] Studies of SPs in noble metals have revealed that light energy can be confined in volumes of deep sub-wavelength dimensions[2] or propagated in metallic stripes[1] and grooves[3] over tens of micrometers. The appealing prospects of merging both properties into integrated structures and so of further pushing the limit of plasmonics towards atomic-scale devices face significant challenges in high fidelity fabrication and sub-nanometer-scale characterization of electromagnetic fields. In this regard, crystalline noble metal colloids act as a generic platform to tailor plasmonic properties in extremely small volumes near well-defined crystallographic surfaces, which has opened the way to efficient sub-wavelength propagation with significantly reduced dissipation, nanometer-scale confinement of the electromagnetic field and quantum plasmonics.[4-6] SP modes in individual SP-bearing nanoparticles (NPs) can be designed by controlling their morphology, such as crystalline 1D nanowires,[7] 2D ultrathin platelets [8,9] or 3D dendritic NPs, [10] which directly shapes spatially and spectrally the SP local density of states (SP-LDOS). Alternatively, new SP modes emerge by tuning the inter-particle dipolar SP coupling in dimers or linear arrays.[11] More complex assemblies are conveniently produced by self-assembling small nanoparticles[12] into superstructures such as chains[13] and chain networks,[14,15] 2D sheets[16] or 3D assemblies.[17,18] The small particle size and interparticle gap lead to large electromagnetic field enhancement yet long-range delocalization is intrinsically limited by the capacitive nature of the SP coupling.[11] The control of the conductive connectivity of possibly crystalline building blocks appears as a challenge of paramount importance for higher order plasmonic architectures.[19,20]

In this Article, we show that ultimately narrow waveguiding structures are obtained by converting the capacitively coupled SPs in assemblies of strongly interacting NPs into conductively delocalized SP modes in continuous crystalline structures that retain the native morphology of the complex self-assembled blueprint. We prepare fused metallic bead strings by electron beam-induced interparticle melting of self-assembled 10-nm Au nanoparticle chain networks. In such networks of ultrathin chains, mapping the plasmon field and theoretically predicted SP-LDOS confinement[21] is an experimental challenge. Here, we use monochromated electron energy loss spectroscopy (EELS) to reveal the spatial and spectral distribution of the SP-LDOS that fully characterize the modal behavior of our complex fused nanoparticle networks. [8,22-26] We demonstrate that the fused superstructures tailor the SP-LDOS with an unprecedented spatial resolution and a convenient spectral addressability. The SP-LDOS is shown to be tightly confined within a few tens of nanometers around the NPs along highly contrasted and energy-dependent hot spot patterns that suggest a nanoscale spatial and spectral input/output addressability of light energy in these fused chain networks. Moreover, we observe the emergence of SP modes with energy as low as 0.38 eV (3200 nm equivalent photon wavelength) that are suitable for long range propagation in nanoscale-wide waveguides, near-IR sensing, light energy up-conversion or localized heat sources. These findings of SP-LDOS design for passive information transport echo new concepts that have been put forward for active optical information processing using colloids, such as logic gate devices based on interferential engineering in coupled colloids [27] or on the design of the SP local density of states (SP-LDOS) in large crystalline colloids.[9] Our results suggest that the fused chains are ideal objects to explore crystalline plasmonic circuits at the limit between classical and quantum behavior.

Monodisperse gold nanoparticles, 12 nm in diameter, are prepared by the standard citrate reduction of gold tetrachloroaurate(III) and undergo a spontaneous self-assembly into complex yet well-reproduced extended plasmonic nanoparticle networks (PNNs) upon addition of an adjusted amount of mercaptoethanol (MEA) following the protocol detailed in reference [14] and summarized in Methods.[28] PNNs are micrometer-sized reticulated networks of single-particle chains comprising typically 10-20 crystalline nanoparticles between nodes and thousands of particles overall and a 1-nm interparticle gap filled with MEA (Fig. S1). For EELS experiments, the PNNs are deposited on ultrathin amorphous $SiN_x$ membranes, briefly exposed to an $O_2$ plasma in order to remove the MEA capping layer, and rapidly introduced into the transmission electron microscope (TEM). Significantly, in the absence of the $O_2$ plasma treatment, the PNNs are stable under an 80-100kV electron beam allowing detailed structural characterization (Figs. S1 and S2A). However, after the plasma cleaning, a near instantaneous fusion of the metal is selectively triggered in the narrow interparticle gap regions (Figs. 1A-B). The native self-assembled PNNs are thus converted *in-situ* into continuous and crystalline metallic bead strings that retain the network morphology (Fig. S2). We ascribe the localized melting of metallic nanoparticles to an enhanced local electric field in gap regions induced by the electron irradiation.[29,30]

EELS maps of fused PNNs are recorded in scanning TEM mode at 80 kV by using a modified Spectrum Imaging technique [31] detailed in Methods.[28] After data processing, 0.1 eV energy windows around the plasmon peaks of interest are used to map the EELS intensity in each pixel. In Figure 1C, the local spectra of two specific locations, I and II, near the fused particle loop shown in Fig. 1B, are compared to the bulk extinction spectrum of the pristine, non-fused, PNNs in aqueous suspension (black dotted line). All three spectra share a low intensity peak located at 2.45 eV (i.e. 520 nm), characteristic of the transverse SP mode of single particle chains and originating in the SP mode of the isolated 12-nm Au nanoparticles.[14] The uniaxial coupling of the localized SPs in the non-fused chains results in the emergence of lower energy modes that are inhomogeneously combined into the broad peak centered at 1.85 eV of the extinction spectrum. This longitudinal peak is accounted for by considering an average 1 nm interparticle gap filled with MEA and citrate molecules of average index 1.6.[14] Interestingly, after fusion, the EELS spectra shown in Fig. 1C also present a second low energy peak, although its maximum is markedly red-shifted to 1.61 (position I) or 1.11 eV (position II). The corresponding EELS maps in Figs. 1C-E illustrate the spatial distribution of these three SP modes along the particle loop. The transverse mode (2.45 eV) appears to be strongly and uniformly confined along the entire contour (Fig. 1D). In contrast, the longitudinal modes exhibit a more pronounced modulation with the 1.61 eV resonance concentrated across the vertical diameter of the loop and the 1.11 eV mode predominantly distributed along the horizontal diameter with a maximum at the tip of the chain. These spatial modulations are reminiscent of other higher order longitudinal SP modes observed by near-field probe techniques in nanowires.[4,7,32]

In order to get a better insight into the spectral and spatial evolution of the SP modes in the fused PNN, we have developed a model and a numerical tool dedicated to the simulation of EELS experiments and based on the Green Dyadic Method (GDM). We consider a swift electron beam probing the near-field of a linear nanoparticle chain in normal incidence at location $R_0$ (Fig. 2A). In such a configuration, EELS maps and spectra can be computed from the LDOS projected onto the electron trajectory (Z axis) by calculating the total Coulomb force work accumulated by the electron.[33] We formulate an equivalent quantum approach in which the impinging electrons that interact with the plasmonic structure undergo a state change and yield a $\hbar\omega_0$ quantum of energy. This transition generates an effective dipole oscillation at the loss frequency $\omega_0$. Since its field linearly derives from the Green dyadic response of the nanoparticle, a formal relation between the average energy lost per electron and per unit time, $\Delta E_{EELS}^z$, and the imaginary part of the (zz) component of the Green dyadic tensor, $S_{zz}(\mathbf{R_0}, \mathbf{R_0}, \omega_0)$, can be easily established:[28]

$$\Delta E_{EELS}^z(\omega_0) = \frac{e^2\hbar^2\mathcal{A}^2}{8m^2\omega_0}\Im\{S_{zz}(\mathbf{R_0},\mathbf{R_0},\omega_0)\} \qquad (1)$$

where *m* and *e* are the electronic mass and charge, $\mathcal{A}$ accounts for the electron probe resolution and has the dimension of an inverse length and $\omega_0$ is the angular frequency of the energy loss.

Therefore, the knowledge of the field propagator $S_{zz}(\boldsymbol{R_0}, \boldsymbol{R_0}, \omega_0)$ associated with any nanoparticle superstructure suffices to compute the EELS maps and spectra. In general, numerical simulations are performed on finely discretized structures as shown in Figure 2. We first examine the spectral evolution of the EELS signal of a linear chain comprising an increasing number of fused nanoparticles (Fig. 2C). For a dimer of fused particles, two peaks are visible. The high energy transverse mode (2.40 eV) corresponds to the SP mode of the isolated particle (2.42 eV) that has undergone a marked attenuation and a small red-shift. The energy position of this transverse mode remains essentially unchanged as the number of fused particles increases, whereas the attenuation proceeds monotonously. Noticeably, the low energy peak rapidly shifts from 2.10 eV, for the dimer, towards lower energy, without significant broadening. This is very different from the case of self-assembled, non-fused particle chains for which the low energy resonance rapidly saturates at a value that depends on the gap size and nature of the capping layer.[15] For MEA-capped PNNs, the limit is ca. 700 nm for an average gap size of 1 nm (Fig. 1C).[14] The observed linear increase of the wavelength of the peak maximum with the aspect ratio of the fused chain (Fig. 2C, inset) is similar to the classic behavior of Au nanorods (Fig. S5).[34] This result strongly suggests that the large red-shift observed in the EELS spectra of Fig. 1C, but not in the absorption spectrum of native PNNs, can be ascribed to the local fusion of neighboring particles. Moderate fusion between nanoparticles produces enough metallic contact to restore topological order and to open new conductive SP channels. The resulting marked low-energy shift of the modes indicates an increased physical size of the supported coherent excitations. A similar influence of conductive coupling was observed in nanorods or tip-to-tip triangular prism dimers upon breaking the metallic bridge.[19,20] The calculation of the plasmonic transmittance though fused nanoparticle chains (Fig. S3) further underpins the efficient guiding of optical information in fused PNNs. It is noteworthy that the fused bead strings sustain fewer and broader SP modes than the sharp resonator behavior of perfectly straight and smooth nanorods because of the residual surface corrugation and general morphological complexity.

Applying our simulation tool to more realistic structures, we constructed a model of the sample in Fig. 1B, where each particle is assimilated to a sphere fused with its nearest neighbors according to the projection given by the TEM image. The entire structure volume is discretized on a face centered cubic mesh in order to better account for the bead string morphology (Fig 2B). The agreement between the experimental EELS (Figs. 1D-F) and simulations maps (Figs. 1G-I) is remarkable. In particular, the homogeneous contour of the 2.45 eV map and the strong confinement of the EELS signal within 15 nm from the loop contour can be observed in both experimental (Fig. 1D) and simulated (Fig. 1G) maps. Equivalently, the overall EELS signal distribution as well as the specific localized intensity maxima of the 1.61 eV (Figs. 1E and 1H) and 1.11 eV (Figs. 1F and 1I) maps coincide. The simulated EELS spectra present three characteristic peaks at 2.37 eV (Fig. 1J), 1.67 eV (Fig. 1K) and 1.04 eV (Fig. 1L) in close match with the recorded data.

Figure 3A features a larger PNN fragment comprising several loops and chains of fused nanoparticles. The EELS map of the transverse SP mode recorded at 2.40 eV (Fig. 3B) confirms its extreme and homogeneous confinement along the edge of the entire structure, in agreement with earlier calculations.[21] Since this high energy mode is impervious to the effect of the limited fusion of the particle chains, we computed the EELS map shown in Fig. 3C by describing each sphere with a dipolar polarizability.[21] Despite this approximation, the simulated map reproduces the experimental data with fine details, including the areas of higher EELS intensity. These maps can be more quantitatively analyzed by comparing two intensity profiles extracted along the same line (Fig. 3D). The experimental EELS intensity (blue histogram) and the simulated signal (continuous orange line) overlap exactly. In particular the decay of the EELS signal away from the particle edges is accurately described. Figures 3E and 3F accumulate cross-sections of the near-field decay range taken horizontally inside the black box of Fig. 3A and with the distance origin chosen at the particle edge. Both experimental and simulated decay curves confirm that the confinement of the LDOS at this energy is on the order of 15 nm. Our results suggest that the decay rate follows a slower trend than a $r^{-3}$ dipolar power law profile that could be expected from the classical description of the near-field near the surface of a single nanoparticle,[35] for example in fluorescence experiments.[36] Among possible causes of this variation, the complex morphology of the fused PNN places any point in the near-field under the influence of a large number of particles and the EELS signal in a given position accumulates the energy loss along the entire electron trajectory rather than in a single point-like location.

With small bead strings sustaining modes with energy as low as 1 eV (1240 nm), we anticipate that larger fused PNNs hold a strong potential for long-range propagation of ultimately confined, low energy SP modes. Fig. 4A shows EELS spectra recorded down to 0.2 eV in locations I to IV near the fused PNN of Fig. 3A. The ubiquitous 2.45 eV transverse resonance is weak in comparison to the multimodal features observed between 1.5 and 0.2 eV, the spectral details of which vary from one location to another. Linear or kinked chains in locations II and III yield essentially a single peak around 0.7-0.9 eV, while branched and looped topologies in locations I and IV sustain a multiple peak pattern with dominant features 0.4 and 1.2 eV. A systematic analysis of EELS spectra associated with basic topological patterns confirms that Y-shaped junctions, free-end chains and linear fragments produce specific SP-LDOS resonances around 0.4, 0.9 and 1.5 eV (Figs. S6-S8).[28] Remarkably, similar findings were reported for calculated far-field extinction spectra of linear, kinked, branched and looped fragments of aqueous PNN suspensions.[14] The multiple resonances of the longitudinally coupled SP mode occurred at higher energies (1.55-1.8 eV) since the chains were not fused but could nevertheless be observed due to the large local refractive index created by the organic capping layer.[28] The spectral features in the longitudinal SP modes are therefore, in part, determined by the topology that pre-exists in the native PNN and are preserved in the fused bead strings.

Interfacial fusion further enhances the coupling between particles by creating continuous, conductive structures that lead to the long range delocalization indicated by the extremely low energy modes probed in EELS. The spatial extension of each of these modes was characterized by mapping the EELS intensity as presented in Figs. 4B-F (see also Fig. S7). The well-resolved maps reveal that the EELS intensity or, equivalently, the SP-LDOS at the chosen energy is tightly confined near the narrow particle chains. Moreover, the intensity extrema present marked spatial variations when the 0.1 eV-wide energy window is tuned between 1.2 and 0.4 eV. The transverse confinement naturally loosens as lower energies are probed. The apparent transverse decay length of the EELS signal is *ca.* 20 nm at 1.21 and 1.00 eV (Fig. 4B and 4C) but reaches *ca.* 50 nm at 0.60 and 0.38 eV (Fig. 4E and 4F). The multimodal behavior of the fused PNN creates a collection of high intensity spots, 30 to 50 nm in diameter, that are located at small loops (Figs. 4B, right side or 4E top left)), at successive spots along the central chain (Figs. 4F, 4C, 4D, 4E), inside the large loop (Figs. 4B, 4D, 4F) or at the very end of lateral chains (Fig. 4E). The selective spectral tuning of the spatial distribution of SP-LDOS in PNN, which was predicted for visible wavelengths in pristine networks,[21] can be extended, after fusion, to the mid-IR range.

Interestingly, in some cases, an EELS signal that continuously follows the colloidal superstructure is measured for some energies (for example 1.21, 1.00 and, to a lesser extent, 0.38 eV) while other energy windows clearly segregate the network into resonant and non-resonant areas (for example 0.60 and 0.79 eV). Optically, the fused structures can be considered as 12-nm wide SP waveguides with complex topology in which some SP modes promote energy propagation along metallic bead strings, when other ones confine it in specific sub-20 nm fragments. Indeed, GDM near-field calculations presented in Fig. S3 confirm the significant transmittance of fused NP chains upon excitation by a dipole source placed near the proximal end of the chain. The transmittance spectrum exhibits clear modal features that reflect the SP-LDOS spectrum. When excited at the wavelength of a low-energy resonance maximum, the transmitted near-field intensity at 10 nm above and away from the distal end can exceed 70% of the intensity at the input. Although the PNN fragments displayed are small, micrometer-long examples show similar properties (See Fig. S9) and the extent of fully formed PNN can reach a globular size of several micrometers in diameter opening the way to meso- to nanoscale interfacing. In this context, regions where the EELS signal shows a maximum (*i.e.* large SP-LDOS) can be conceived as entry points to address the waveguiding networks optically [37] or inelastically using a low bias tunneling current.[38,39]

In conclusion, this work demonstrates sculpting SP-LDOS with unprecedented spatial resolution and a convenient spectral addressability. In particular, fused PNNs gather several desirable attributes for ultimate scale optical applications. The self-assembled PNNs are topological blueprints that can be subsequently fused into multimodal plasmonic waveguides offering nanoscale lateral confinement and micrometer scale transport tracks. The initial self-assembly step sets the spatial SP-LDOS landscape by defining the topology of the nanoparticle ensemble. Next, the local fusion spectrally converts the coupled local SP modes into extended SP channels and reduces the overall disorder. Our approach is general and can be applied to the fusion of better ordered superstructures. Indeed, the complexity of self-assembled PNN could be harnessed into more regular constructs by precisely directing the self-organization of crystalline metal building blocks in lithographically designed templates.[18,40]

The wealth of high and low energy SP modes bound in the fused chain networks opens the possibility of spectral addressing and spatial control of light energy at the nanometer length scale that can be exploited in a number of areas. The strong and addressable confinement of the near-IR resonances could trigger highly localized surface-enhanced IR absorbance (SEIRA) and emission of single near-IR fluorophores such as lanthanide-containing molecules [41] or polyaromatic hydrocarbons[42] and thus contribute to a more efficient up-conversion of light energy in photovoltaic devices. They could equally improve the interfacing of other near-IR active two-dimensional materials such as graphene nanoribbons.[43,44] Complex colloidal architectures such as PNNs or their future templated derivatives could also be engineered as localized heat sources [45] or metamaterials [17] for which the fusion post-treatment could advantageously tune the SP modes to reinforce the field enhancement or modulate the refractive index at chosen wavelengths.

Finally, PNNs are ideal systems to explore the limit between classical and quantum plasmonics in an extended ensemble of 10-nm nanoparticles spaced by sub-1 nm gaps that can be further decreased upon controlled fusion.[46,47] We show that colloidal self-assembly fosters a promising route to demonstrate concepts of quantum plasmonic circuitry that might even be driven down to atomic scale electro-optical addressing in atomic metal chains.[48]

**Methods**

*Nanoparticle chain self-assembly:* Plasmonic Nanoparticle Networks (PNN) were synthesized by the method reported in references [14,49]. Briefly, Au nanoparticles were freshly prepared by the citrate reduction method at a citrate : $[AuCl_4]^-$ molar ratio of 5.2 : 1 and diluted to the required concentration with 18 MΩ.cm deionized water. The average diameter of the Au nanoparticles was 12.0 ± 1.1 nm. The assembly of the PNN was performed at room temperature by adding 2-mercaptoethanol ($HS(CH_2)_2OH$) to the diluted Au nanoparticle solution at a Au nanoparticles : MEA molar ratio of 1 : 5000. The nanoparticle chain assembly is characterized by a color change from pink to purple as the coupled modes emerge. It was monitored by UV-Visible spectrophotometry until completion within 24 to 48 h after mixing.

*TEM sample preparation:* A 10 µL droplet of fully formed PNN suspension was drop-casted onto 10-nm thick silicon nitride membranes and left to dry in clean environment. A series of 1-minute $O_2$ plasma cleaning steps were performed to eliminate the mercaptoethanol and citrate capping moieties. Structural TEM analysis was performed using a Philips CM20FEG microscope operated at 100 kV. Particular care was taken to adjust the plasma cleaning step to avoid any damage or modification of the chain morphology compared to unprocessed PNN. Complementary SEM observations were performed on a Zeiss 1540XB Gemini microscope.

*Electron energy-loss spectroscopy (EELS):* EELS was performed in scanning TEM (STEM) mode using an FEI Titan TEM with Schottky electron source. The microscope was operated at 80 kV, and a STEM convergence semi-angle of 13 mrad was used to form a probe with a diameter of approximately 1 nm. A Wien-type monochromator dispersed the electron beam in energy, and an energy-selecting slit formed a monochrome electron beam with typical full-width at half-maximum values of 70 meV. A Gatan Tridiem ER EELS detector was used for EELS mapping and spectroscopy, applying a 12 mrad collection semi-angle. EELS data was acquired with a modified binned gain averaging routine:[31] individual spectra were acquired in 40 ms, using 8 or 16 times on-chip binning. The detector channel-to-channel gain variation was averaged out by constantly changing the readout location and correcting for these shifts after the EELS acquisition was finished. A high-quality dark reference was acquired separately, and used for post-acquisition dark signal correction. Spectra were normalized by giving the maximum of the zero-loss peak (ZLP) unit value, and the ZLP background signal was removed by fitting and subtracting a high-quality background spectrum.

***EELS mapping:*** EELS maps were obtained with the Spectrum Imaging technique: scanning a small electron probe with an approximate diameter of 1 nm in a rectangular raster of pixels, while at each pixel an EELS spectrum is collected and stored. After data processing as described above, 0.1 eV energy windows around the plasmon peaks of interest were used to image the EELS intensity in each pixel in linear scale. The EELS intensity maps were colour-coded to a temperature scale.

***Simulations:*** Our model formulates the energy yielded by a swift electron passing in the vicinity of a metallic object in terms of an effective dipole oscillating at the loss frequency. The average power transferred to the plasmonic system is computed by the Green Dyadic Method in which the whole dyad $S(\mathbf{r},\mathbf{r'},\omega_0)$ of the considered system is computed by solving a Dyson's equation sequence.[50] In order to optimize the representation of the complex curved surfaces of the fused bead chains, the volume of the system is discretized in a face centered cubic lattice. This method allows the numerical computation of EELS maps for a given energy, $\hbar\omega_0$, and spectra for a fixed position, $\boldsymbol{R_0}$.

**Acknowledgments**

The authors thank Prof. A. Mlayah and Dr. A. Arbouet for continuous and fruitful discussions. This work was supported by the European Research Council (ERC) (contract number ERC–2007-StG Nr 203872 COMOSYEL), the massively parallel computing center CALMIP in Toulouse. A. T. thanks the LabEx project NEXT for a travel grant to IMRE. The authors thank M. Nunez for technical assistance in TEM imaging.

**Author contributions**

E.D., M.B., C. G. and S. M. conceived the experiments. G. K. L. and M.L. synthesized the nanoparticles and A.T. and M. L. prepared PNN suspension and TEM samples. M.B. performed and processed EELS experiments. C.G. and A.T. developed the model and implemented the simulation codes. A.T., M.B., C.G. and E.D performed data and simulation analysis and prepared the figures. All co-authors contributed to the writing of the article.

**Additional information.**

Supplementary information is available in the online version of the paper. (Materials and Methods. Figures S1-S9. Theoretical model and numerical simulation of EELS signal. Statistical correlation between SP modes and topological features. Principal Component Analysis.).

Correspondence and requests for materials should be addressed to E.D. or M. B.

**Competing financial interests**

The authors declare no competing financial interests.

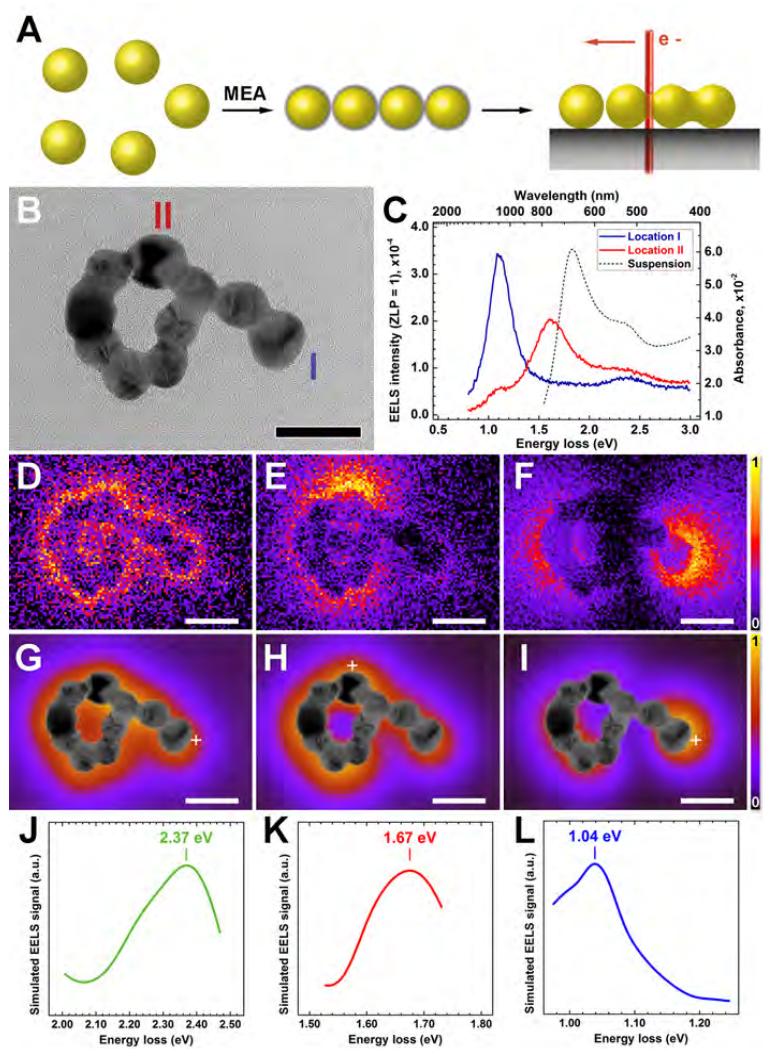

**Fig. 1. Spatial and spectral characterization of plasmon-mediated electron energy-loss in the vicinity of a fused Au nanoparticle chain.** (**A**) Scheme of mercaptoethanol-driven PNNs self-assembly and their conversion into continuous bead strings by electron-beam induced welding. (**B**) TEM image of a short looped Au particle chain after local fusion. The individual particles still display their penta-twinned structure. Markers (I) and (II) indicate the position where loss spectra displayed in (C) are recorded. (**C**) EELS spectra recorded in position (I), blue, and (II), red. The absorption spectrum of the starting particle networks suspended in water is displayed by the black dotted line. The equivalent photon wavelength axis is computed as λ(nm) = 1240/Energy(eV). (**D-F**) Experimental and (**G-I**) simulated EELS maps recorded at (**D, G**) 2.45 eV, (**E, H**) 1.61 eV and (**F, I**) 1.11 eV. In (**G-I**), the maps are calculated in a plane 10 nm above the nanoparticles and the TEM outline of the fused particle chain is overlaid. (**J-L**) Calculated EELS spectra obtained from equation (1) at locations close to (I) and (II) and cross-marked in (**G-I**) respectively. All scale bars are 20 nm.

A. Teulle et al.

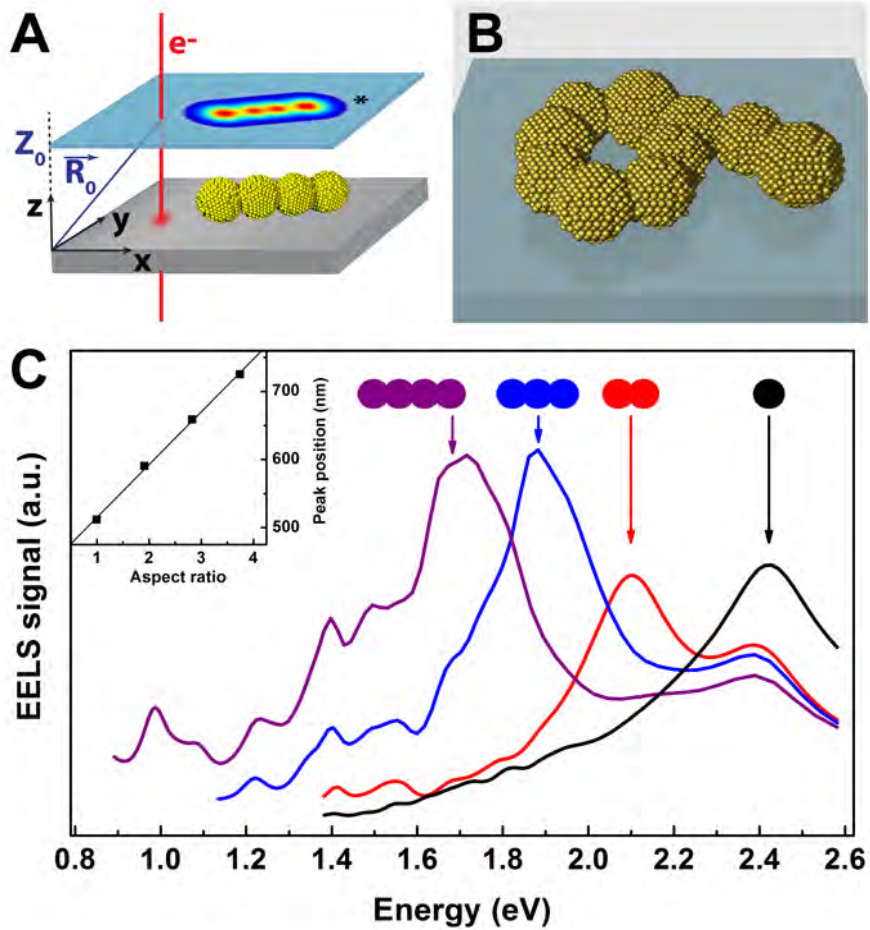

**Fig. 2. Numerical simulation of EELS spectra of plasmonic nanoparticle chains.** (**A**) Schematic of a swift electron beam impinging at $R_0 = (X_0, Y_0, Z_0)$ to probe the near-field of a linear nanoparticle chain deposited on a thin membrane. The inelastic energy loss process is computed in a point located along the chain axis at a distance $2a$ from the terminal particle center in the plane $Z_0 = 2a + 10\,\text{nm}$ with "a" the nanoparticle radius (see star marker). (**B**) Schematic of the discretized face centered cubic 3D model of the fused particle chain shown in Fig. 1B. (**C**) Simulated EELS spectra of an isolated 12-nm Au nanoparticle (black) and a series of fused linear chains composed of two (red), three (blue) and four (purple) Au nanoparticles. The fused overlap corresponds to 17% of the diameter. Inset: Linear evolution of the low energy peak position as a function of the fused chain aspect ratio, $L_{chain} / 2a$, with $L_{chain}$ the total chain length. The continuous line is a linear fit to the data.

A. Teulle et al.

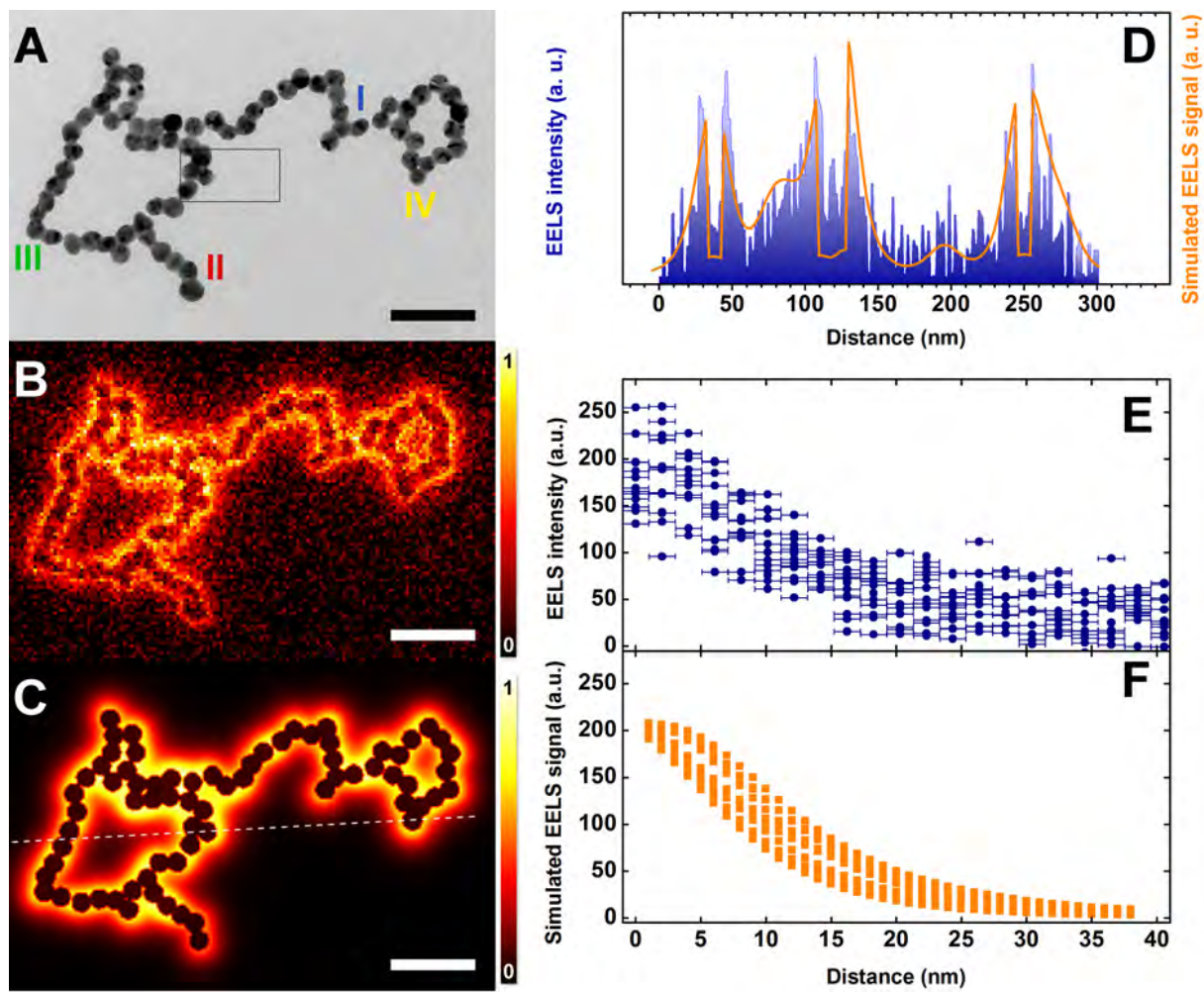

**Fig. 3. Extreme confinement of the 2.4 eV plasmon mode in complex PNN.** (**A**) TEM image of a branched and looped Au nanoparticle chain after undergoing in-situ electron beam induced fusion. (**B**) EELS intensity map recorded at 2.4 eV corresponding to the maximum of the transverse plasmon mode. (**C**) Simulated EELS map at 2.4 eV computed in the DDA approximation. The elastic electron scattering is taken into account by nulling the EELS signal in the position of the nanoparticles. Scale bars are 50 nm. (**D**) Experimental (blue histogram) and simulated (orange line) single line cross-sections extracted from (B) and (C), respectively, along the dotted line shown in (C). (**E**) Experimental and (**F**) simulated spatial decay of the EELS intensity extracted from horizontal cross sections in maps (B) and (C) respectively in the region marked by the thin black box in (A).

A. Teulle et al.

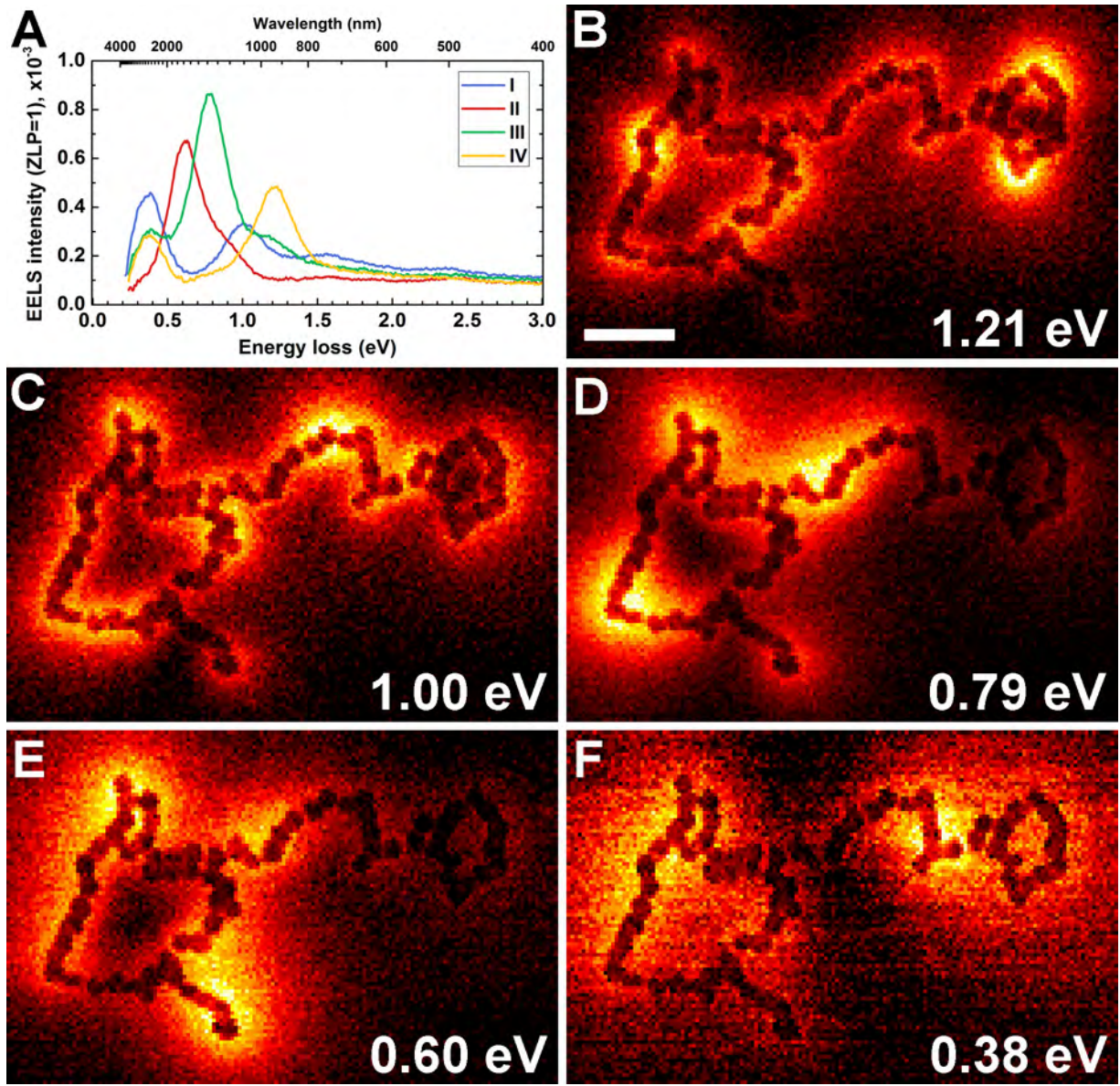

**Fig. 4. Mapping of low energy plasmon modes in complex PNN**. (A) EELS spectra recorded in positions (I) to (IV) near the PNN structure shown in Fig. 3(A). The equivalent photon wavelength axis is computed as λ(nm) = 1240/Energy(eV). (B-F) EELS maps recorded from the well-separated resonance features at 1.21, 1.00, 0.79, 0.6 and 0.38 eV with a 0.1 eV pass band width. Scale bars are 50 nm. Color scale is similar to Fig. 3B.

A. Teulle et al.

# Supplementary Materials

## Multimodal plasmonics in fused colloidal networks

Alexandre Teulle[1], Michel Bosman[2]*, Christian Girard[1], Kargal L. Gurunatha[1], Mei Li[3], Stephen Mann[3], Erik Dujardin[1]*

[1] CEMES CNRS UPR 8011, 29 rue J. Marvig, 31055 Toulouse Cedex 4, France.

[2] Institute of Materials Research and Engineering, A*STAR (Agency for Science, Technology and Research), 3 Research Link, 117602 Singapore.

[3] Centre for Organized Matter Chemistry, School of Chemistry, University of Bristol, Cantocks Close, Bristol BS81TS, United Kingdom.

**Table of Contents**

## 1. Materials and Methods

**Nanoparticle chain self-assembly:** Plasmonic Nanoparticle Networks (PNN) were synthesized by the method reported in references [1,2]. Briefly, Au nanoparticles were freshly prepared by the citrate reduction method at a citrate : $[AuCl_4]^-$ molar ratio of 5.2 : 1 and diluted to the required concentration with 18 MΩ.cm deionized water. The average diameter of the Au nanoparticles was 12.0 ± 1.1 nm. The assembly of the PNN was performed at room temperature by adding 2-mercaptoethanol ($HS(CH_2)_2OH$) to the diluted Au nanoparticle solution at a Au nanoparticles : MEA molar ratio of 1 : 5000. The nanoparticle chain assembly is characterized by a color change from pink to purple as the coupled modes emerge. It was monitored by UV-Visible spectrophotometry until completion within 24 to 48 h after mixing.

**TEM sample preparation:** A 10 μL droplet of fully formed PNN suspension was drop-casted onto 10-nm thick silicon nitride membranes and left to dry in clean environment. A series of 1-minute $O_2$ plasma cleaning steps were performed to eliminate the mercaptoethanol and citrate capping moieties. Structural TEM analysis was performed using a Philips CM20FEG microscope operated at 100 kV. Particular care was taken to adjust the plasma cleaning step to avoid any damage or modification of the chain morphology compared to unprocessed PNN. Complementary SEM observations were performed on a Zeiss 1540XB Gemini microscope.

**Electron energy-loss spectroscopy (EELS):** EELS was performed in scanning TEM (STEM) mode using an FEI Titan TEM with Schottky electron source. The microscope was operated at 80 kV, and a STEM convergence semi-angle of 13 mrad was used to form a probe with a diameter of approximately 1 nm. A Wien-type monochromator dispersed the electron beam in energy, and an energy-selecting slit formed a monochrome electron beam with typical full-width at half-maximum values of 70 meV. A Gatan Tridiem ER EELS detector was used for EELS mapping and spectroscopy, applying a 12 mrad collection semi-angle. EELS data was acquired with a modified binned gain averaging routine:[3] individual spectra were acquired in 40 ms, using 8 or 16 times on-chip binning. The detector channel-to-channel gain variation was averaged out by constantly changing the readout location and correcting for these shifts after the EELS acquisition was finished. A high-quality dark reference was acquired separately, and used for post-acquisition dark signal correction. Spectra were normalized by giving the maximum of the zero-loss peak (ZLP) unit value, and the ZLP background signal was removed by fitting and subtracting a high-quality background spectrum.

**EELS mapping:** EELS maps were obtained with the Spectrum Imaging technique: scanning a small electron probe with an approximate diameter of 1 nm in a rectangular raster of pixels, while at each pixel an EELS spectrum is collected and stored. After data processing as described above, 0.1 eV energy

A. Teulle et al. 2014

windows around the plasmon peaks of interest were used to image the EELS intensity in each pixel in linear scale. The EELS intensity maps were colour-coded to a temperature scale.

**Simulations:** Our model formulates the energy yielded by a swift electron passing in the vicinity of a metallic object in terms of an effective dipole oscillating at the loss frequency. The average power transferred to the plasmonic system is computed by the Green Dyadic Method in which the whole dyad $S(\mathbf{r},\mathbf{r}',\omega_0)$ of the considered system is calculated by solving a Dyson's equation sequence.[4] In order to optimize the representation of the complex curved surfaces of the fused bead chains, the volume of the system is discretized in a face centered cubic lattice. This method allows the numerical computation of EELS maps in planes located 10 nm above the top of the nanoparticles for a given energy, $\hbar\omega_0$, and spectra for a fixed position, $\boldsymbol{R_0} = (X_0, Y_0, Z_0)$.

NOTE: This method is closely related to the one used to describe an optical excitation of plasmonic nanostructures such as PNN which we described in an earlier work.[5] In this case other parameter can be explored such as the polarization of the exciting field

## 2. TEM image of a typical self-assembled PNN network.

The self-assembly driven by the addition of mercaptoethanol (MEA) to citrate-stabilized Au nanoparticles leads to large (2-3 micrometer diameter) globular networks of single particle chains (PNN, Plasmonic Nanoparticle Networks) that can be deposited onto various substrates including TEM grids. The combined collapse and fragmentation of the PNN can be controlled by substrate treatments [6] and can leads to flat spread chain networks of varying particle density as can be seen in Figure S1. In this work, specific attention was paid to rather small PNN fragments for EELS spectroscopy and mapping. Example of such small broken-off fragments can be seen on the periphery of Fig. S1A.

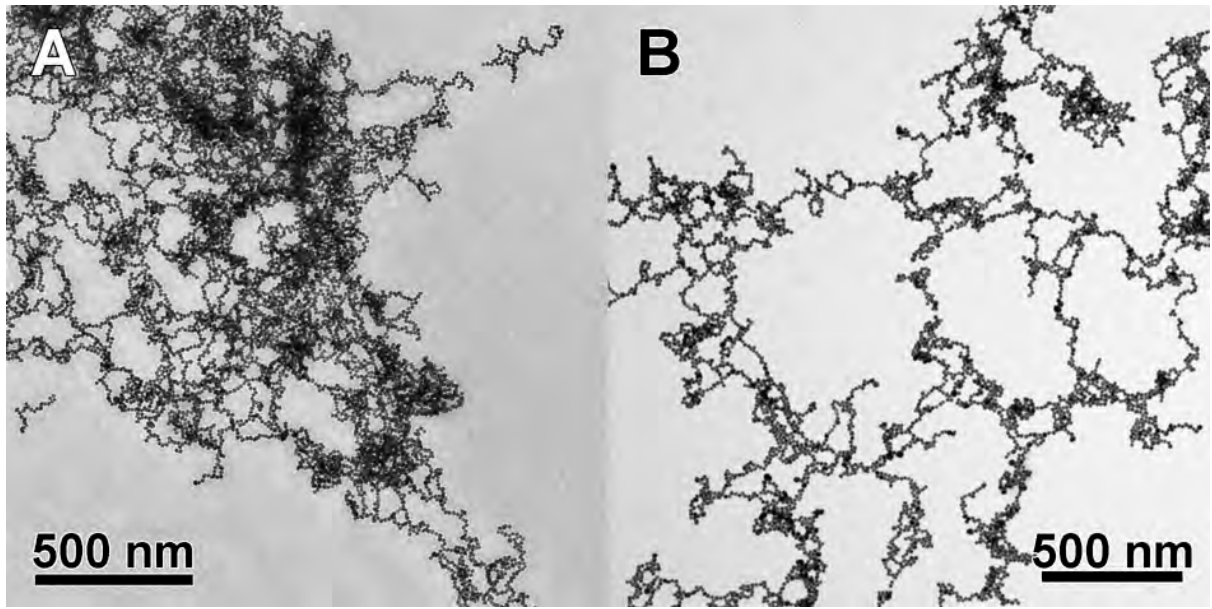

**Figure S1:** (A, B) Low magnification TEM images of portions of full PNN deposited on carbon film.

## 3. Fused plasmonic bead strings by e-beam induced welding of PNN

When PNN are deposited on TEM grids with a formvar / carbon film or on SiN or $SiO_2$ membranes and imaged at 80-120 kV, no significant alteration of the nanoparticles is observed and ample time is allowed to perform structural analysis of the networks as illustrated in Fig. S2A. The self-assembled nanoparticles are initially separated by a 1 nm gap filled with organic molecules (MEA and citrate) that hinder the electron-beam-induced intergrowth.

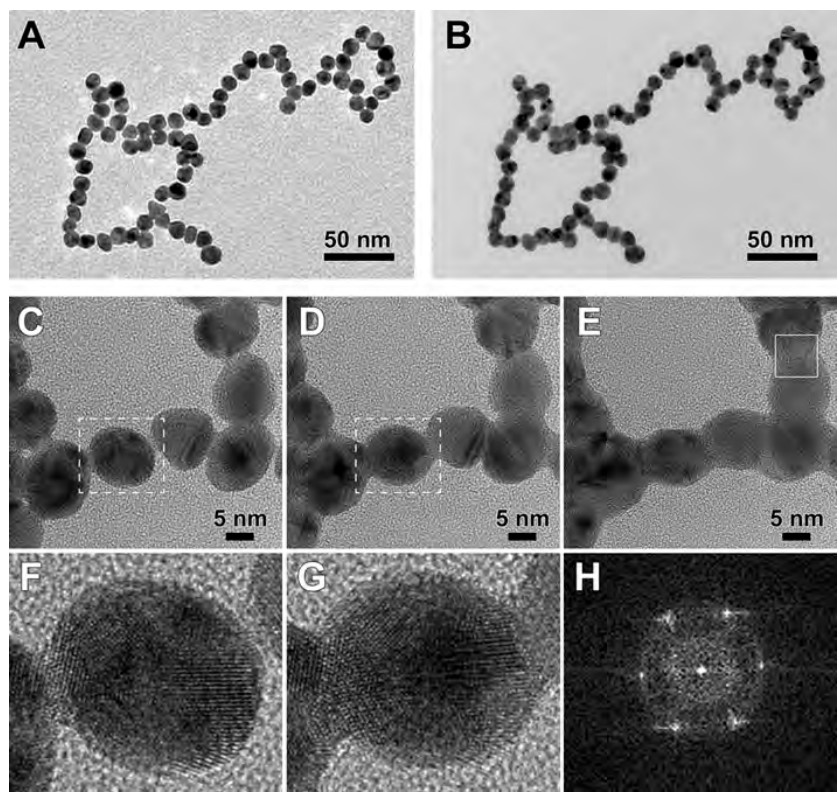

**Figure S2.** Electron beam induced fusion of PNN. (A, B) TEM images of a PNN fragment (A) before $O_2$ plasma cleaning at 80 kV and (B) after $O_2$ plasma cleaning and a few seconds imaging at 200 kV. (C-E) Time-lapsed TEM imaging showing the beam induced fusion of the interfacial areas but preserving the nanoparticle crystallinity. The time intervals are (C-D) 1.3 minutes and (D-E) 4.0 minutes. (F, G) Detail from areas framed in (C, D) showing the crystallinity of the metal bridges formed between neighboring particles during the fusion. (H) Digital Fourier filtering of the interparticle area boxed in (E) after extensive exposure to the electron beam. It shows the six spot of the first Brillouin zone of crystalline gold.

However, when an $O_2$ plasma is performed in order to remove the capping citrate and MEA molecules prior to TEM imaging at 80 or 200 kV, immediate melting of the interparticle spaces can be observed as in Fig. S2B. After the $O_2$ plasma and with the possible influence of the underlying $SiN_x$ substrates, the bare and crystalline metal surfaces in close contact are more easily prone to fusion.

This electron-beam induced fusion was exploited in this work to convert self-assembled PNN into fused particle strings where most particles are linked to their nearest neighbors by crystalline Au bridges resulting from the interfacial welding. The sequence of TEM images in Figs. S2C-E illustrates how PNN are transformed into continuous metallic bead strings and shows that the extent of fusion can be controlled by the electron dose. Close-ups of the gap regions of Figs. S2C and S2D are shown in Figures S2F and S2G. Clearly, the gaps on either side of the particle are rapidly filled with crystalline gold as further underlined by the hexagonal patterned shown in Fig. S2H obtained by Fast Fourier filtering of the boxed gap region in Fig. S2E.

The series of images D-E-F illustrates how the adjustment of the time exposure to the electron beam enables to tune the extent of the welding and the final morphology of the fused bead strings.

## 4. Plasmonic transmittance of fused nanoparticle chains

In this work, fused PNN are conceived as possible plasmonic waveguides with extremely narrow width (e.g. 10 nm) and propagation length in the multi-micrometer length scale when exploiting the lower energy SP modes identified by EELS. While the experimental testing of this proposal is still beyond the capabilities of any available techniques, we have applied our GDM numerical tools that faithfully account for the EELS spatial and spectral data to the calculation of the plasmonic transmittance through a fused nanoparticle chain.

The model systems consists of a linear chain of 5 or 6 Au nanoparticles (diameter 12 nm) similar to those observed in fused PNN. Fig S3A depicts the input-chain-output system. The chain is excited by an optical dipolar source aligned with the chain and placed 12 nm away from the first particle surface. The spectral distribution of the transmitted near-field intensity is calculated at a distal point aligned with the chain located 12 nm away from the last nanoparticle. Figure S3B shows such spectra for two chains lengths of five (blue) and six (green) fused nanoparticles.

The first striking result is that the transmitted intensity is non-zero, suggesting that fused nanoparticle chains do indeed propagate light in a deep sub-wavelength regime. Moreover, the spectral features of the transmitted light present a modal structure that is influenced by the chain length. The relative weight of these modes tends to favor lower energy modes and the resonances are red shifted as the length increases. For example, see the reduced intensity of the 780, 850 nm peaks and increasing intensities of the 920 and 1050 nm peaks plus the emergence of the 1300 nm peak when increasing from five to six particles.

Near field images can be computed just above the chain surface as shown in the inset of Fig. S3B. When the structure is excited off resonance (ex.: 600 nm), the intensity remains located within the first few nanoparticles. However, when a low energy resonance is excited (ex.: 914 nm), a strong intensity is measured at the distal particle illustrating the delocalization of the modal energy. Interestingly, the transmittance does not require that the energy is continuously distributed along the chain but rather that the modal feature excited at that energy presents some finite presence probability at the chain extremity.

This numerical work further confirms that fused PNN do have the potential of efficiently propagating light energy in a deep sub-wavelength regime.

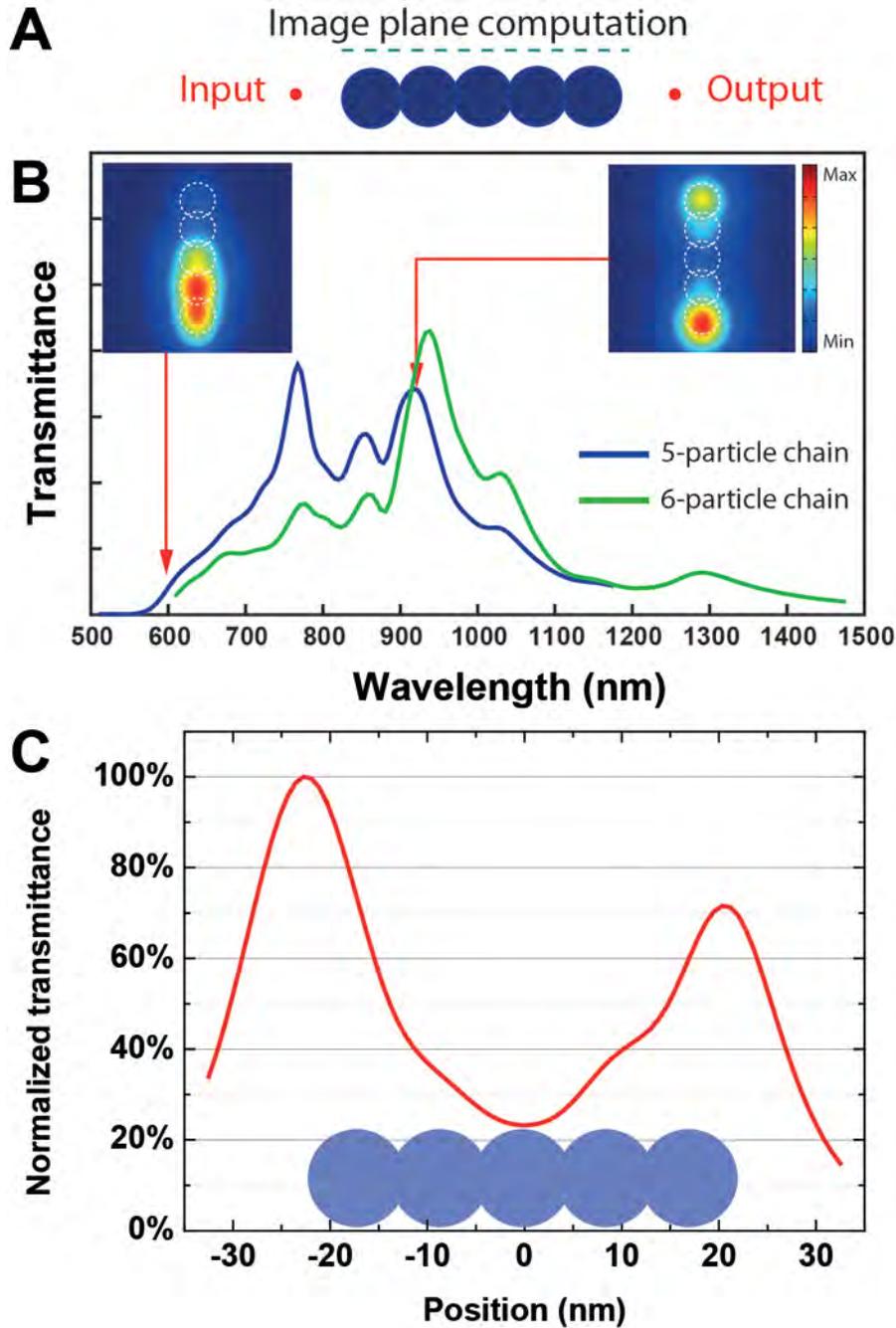

**Figure S3.** Plasmonic transmittance of fused gold particle chains computed from the Green Dyadic Method (GDM). **(A)** Schematics of the input-chain-output system in which the input is excited by an optical dipolar source aligned along the chain axis and the output signal is the near-field intensity generated by the excited plasmonic field. The nanoparticle diameter is 12 nm and the interparticle pitch is 10 nm. **(B)** Transmittance spectra of five- (blue) and six- (green) particle chains. Two near-field optical images have been computed above the 5-particle structure (see the image computation plane depicted in (A)) for two selected excitation wavelengths (off-resonant: 600 nm and resonant: 914 nm). The image size is 65 × 65 nm. (C) Transmission profile along the chain axis of the resonant image (right inset in (B)) that shows a transmitted near-field intensity that exceed 70% of the near-field intensity at the input.[7]

A. Teulle et al. 2014

## 5. Theoretical model and numerical simulation of EELS signals

The theoretical description of EELS phenomena recorded in the vicinity of plasmonic particles can be obtained from the photonic LDOS projected onto the electron trajectory. This statement, demonstrated in reference [8], proceeds from the computation of the total Coulomb force work accumulated during the electron motion.

In this report, we apply an alternative procedure based on a pure Fermi golden rule approach.

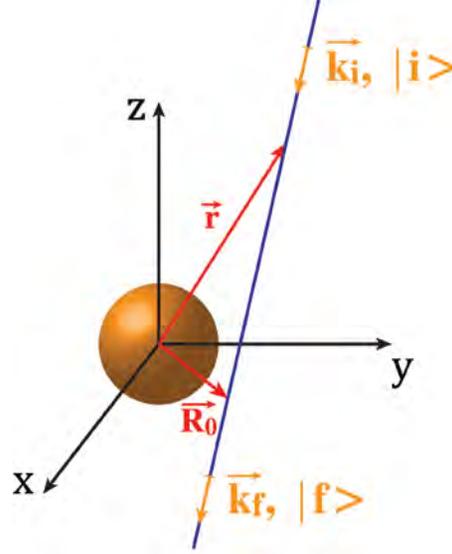

**Figure S4.** Schematic view of a straight electron trajectory passing near a spherical metal particle.

To illustrate how a fast electron can convert a portion $\Delta E_0$ of its kinetic energy into plasmon oscillations, let us consider the schematic drawing depicted in Figure S4. An inelastic event between electron and particle is equivalent to a transition between a first occupied electronic state $|i>$ and a final unoccupied state $|f>$ of the moving electron. The two Bohr frequencies, $\omega_i$ and $\omega_f$, associated with $|i>$ and $|f>$ must verify $\omega_i \geq \omega_f$. The matrix element of the current density associated with this electronic transition is given by the usual relation: [9]

$$\boldsymbol{J}_{i,f}(\boldsymbol{r},t) = -\frac{ie\hbar}{2m}\{\nabla_{\boldsymbol{r}}\psi_f^{\star}(\boldsymbol{r},t)\psi_i(\boldsymbol{r},t) - \psi_f^{\star}(\boldsymbol{r},t)\nabla_{\boldsymbol{r}}\psi_i(\boldsymbol{r},t)\} \quad (1)$$

where $m$ and $e$ are the electron mass and charge, and $\psi_{i/f}$ defines both electronic wavefunctions corresponding to initial and final states $|i>$ and $|f>$, respectively. In the stationary regime corresponding to a constant velocity of the probing electron, before and after the energy transfer event, the electronic wavefunctions display monochromatic oscillation terms:

$$\psi_f(\boldsymbol{r},t) = e^{i\omega_f t}\Phi_f(\boldsymbol{r}) = e^{i(\omega_i - \omega_0)t}\Phi_f(\boldsymbol{r}) \quad (2)$$

and

$$\psi_i(\boldsymbol{r},t) = e^{i\omega_i t}\Phi_i(\boldsymbol{r}) \quad (3)$$

in which, $\omega_0 = \Delta E_0 / \hbar$ is the angular frequency that corresponds to the energy transferred by the swift electron to the plasmonic system and $\Phi_i(\boldsymbol{r})$ and $\Phi_f(\boldsymbol{r})$ define the spatial parts of the electronic wavefunctions, before and after the interaction. After replacing equations (2) and (3) into (1), we can express the real part of the inelastic current as follows:

$$J_{i,f}(r,t) = \frac{e\hbar}{2m} A(r)\cos(\omega_0 t + \Theta(r)), \quad (4)$$

in which, $A(\mathbf{r})$ and $\Theta(\mathbf{r})$ are the modulus and the argument of the complex vector defined by:

$$i\{(\nabla_r)\Phi_f^*(r))\Phi_i(r) - \Phi_f^*(r)(\nabla_r\Phi_i(r))\}. \quad (5)$$

From equation (4), the inelastic current can be expanded in Fourier integral:

$$J_{i,f}(r,t) = \int J_{i,f}(r,\omega)e^{i\omega t}d\omega, \quad (6)$$

with

$$J_{i,f}(r,\omega) = \frac{e\hbar}{2m} A(r)\{\delta(\omega_0 + \omega)e^{i\Theta(r)} + \delta(\omega - \omega_0)e^{-i\Theta(r)}\}, \quad (7)$$

which is associated with the following polarization vector:[10]

$$P_{i,f}(r,\omega) = \frac{ie\hbar}{2m\omega} A(r)\{\delta(\omega_0 + \omega)e^{i\Theta(r)} + \delta(\omega - \omega_0)e^{-i\Theta(r)}\}. \quad (8)$$

According to equation (5), the $A(r)$ vectorial function is built from the product of both electron wavepackets expected before (i) and after (f) the elementary inelastic process occurs. Consequently, it can be schematized by a delta Dirac distribution centered around a location $\mathbf{R}_0 = (X_0, Y_0, Z_0)$ defining the position where the electron released an energy quantum $\hbar\omega_0$:

$$A(r) = A\delta(r - \mathbf{R}_0); \quad (9)$$

in which the modulus $A$ of the vector $A(\mathbf{r})$ has the dimension of an inverse length. This last relation indicates that each inelastic elementary event occurring at the location $\mathbf{R}_0$ of the electron trajectory can be represented by an effective dipole fluctuating at the loss-frequency $\omega_0$:

$$p_{if}(\omega) = \frac{ie\hbar A}{2m\omega}\{\delta(\omega_0 + \omega) + \delta(\omega - \omega_0)\}, \quad (10)$$

in which the phase origin $\Theta(\mathbf{R}_0) = 0$ has been chosen equal to zero for $\mathbf{r} = \mathbf{R}_0$. Now to derive the portion of energy $\Delta E_{EELS}(\omega_0)$, per unit time, transferred by the electron to the metal particle, we just have to apply the laws of electrodynamics of an effective pointlike dipolar light source $p_{if}(t)$, located at $\mathbf{R}_0$. Thus, by applying dynamical Gauss' theorem to $p_{if}(t)$ (cf. reference [10]), we can write:

$$\Delta E_{EELS}(\omega_0) = \frac{\omega_0 p^2}{2} \mathbf{U}\mathbf{U} : \Im \mathbf{S}(\mathbf{R}_0, \mathbf{R}_0, \omega_0), \quad (11)$$

where $p$ is the dipole amplitude defined by:

$$p = \frac{e\hbar A}{2m\omega_0}, \quad (12)$$

$\mathbf{U}$ defines its orientation, and $\mathbf{S}(\mathbf{R}_0, \mathbf{R}_0, \omega_0)$ labels the field-susceptibility of the metal particle. Let us note, that in the particular case where the electron travels on a trajectory parallel to the OZ axis, equation (11) reads:

$$\Delta E_{EELS}^z(\omega_0) = \frac{e^2\hbar^2 A^2}{8m^2\omega_0} \Im\{\mathbf{S}_{zz}(\mathbf{R}_0, \mathbf{R}_0, \omega_0)\}, \quad (13)$$

A. Teulle et al. 2014where $S_{zz}(\mathbf{R_0}, \mathbf{R_0}, \omega_0)$ represents the (zz) component of the dyadic tensor $\mathbf{S}$. The field-propagator $\mathbf{S}(\mathbf{r}, \mathbf{r'}, \omega)$ between two arbitrary points $\mathbf{r}$ and $\mathbf{r'}$ enables to know how the electric field, at location $\mathbf{r}$, is modified near a solid body when a punctual dipolar source is placed in $\mathbf{r'}$.

Two cases can be considered.

**(i)** In the case of a single small metallic particle located at the origin of the framework as depicted in Figure S3, the field propagator $\mathbf{S}(\mathbf{r}, \mathbf{r'}, \omega_0)$ can be composed from the field propagator in vacuum $\mathbf{S_0}(\mathbf{r}, \mathbf{r'}, \omega_0)$ and the dipolar polarizability:[11]

$$\mathbf{S}(\mathbf{r}, \mathbf{r'}, \omega_0) = \mathbf{S_0}(\mathbf{r}, \mathbf{0}, \omega_0) \cdot \alpha(\omega_0) \cdot \mathbf{S_0}(\mathbf{0}, \mathbf{r'}, \omega_0), \quad (14)$$

where $\mathbf{r} = (x, y, z)$ and $\mathbf{r'} = (x', y', z')$. In this latter expression, the vacuum field-susceptibility has been neglected since it does not contribute to the loss of energy.

In a cartesian frame, the $(\beta, \gamma)$ components of $\mathbf{S_0}$ are given by the analytical relation:[10]

$$\mathbf{S}_{0\beta,\gamma}(\mathbf{r}, \mathbf{0}, \omega_0) = \mathbf{S}_{0\beta,\gamma}(\mathbf{0}, \mathbf{r}, \omega_0) = \exp(ik_0 r) \times \frac{3r_\beta r_\gamma - r^2 \delta_{\beta,\gamma}}{r^5}, \quad (15)$$

with $k_0 = \omega_0/c$, ($r_1 = x, r_2 = y, r_3 = z$), and $r = \sqrt{x^2 + y^2 + z^2}$.

In the absence of dispersion spatial effects, the dipolar polarizability of a single metallic sphere of radius $a \ll \lambda_0 = 2\pi c/\omega_0$ placed in vacuum, reads:

$$\alpha(\omega_0) = a^3 \left(\frac{\varepsilon_m(\omega_0) - 1}{\varepsilon_m(\omega_0) + 2}\right), \quad (16)$$

where $\varepsilon_m(\omega_0)$ is the permittivity of the metal.

Expression (13) can be easily applied to an assembly of dipoles using the self consistent scheme based on the discrete dipole approximation (DDA) approach described in ref [11].

**(ii)** For arbitrary shaped metallic particles, the computation of the field-susceptibility $\mathbf{S}(\mathbf{r}, \mathbf{r'}, \omega_0)$ requires a volume discretization of the active region (metal). Thus, the plasmonic system is generally sampled into $n$ identical elementary regions of respective centers $\mathbf{r_j}$ and equal volume $v$, that leads to the following Dyson equation fulfilled by the field-susceptibility $\mathbf{S}(\mathbf{r}, \mathbf{r'}, \omega_0)$:[12]

$$\mathbf{S}(\mathbf{r}, \mathbf{r'}, \omega_0) = \mathbf{S_0}(\mathbf{r}, \mathbf{r'}, \omega_0) + \eta(\omega_0) \sum_{j=1}^{n} \mathbf{S_0}(\mathbf{r}, \mathbf{r_j}, \omega_0) \cdot \mathbf{S}(\mathbf{r_j}, \mathbf{r'}, \omega_0), \quad (17)$$

in which $\eta(\omega_0) = \frac{\varepsilon_m(\omega_0) - 1}{4\pi} v$ is homogeneous to a dipolar polarisability and is directly related to the discretization volume $v$, that itself depends on the discretization grid used to mesh the particles. For a face-centered cubic lattice, this factor reads $v = b^3/\sqrt{(2)}$, where $b$ represents the distance between two consecutive discretization cells. The self-consistent equation (17) is solved using the numerical Dyson's equation sequence procedure detailed in reference [4]. This allows the computation of the field-susceptibility $\mathbf{S}_{zz}(\mathbf{R_0}, \mathbf{R_0}, \omega_0)$ anywhere outside the source region, and the calculation of the EELS signal according to eq (13).

## 6. Compared evolution of peak position with aspect ratio in NP chain, fused bead strings and nanorods

Figure S5 compiles experimental data of the position of (i) the longitudinal SP mode in Au nanorods stabilized by CTAB/DDAB (diameter 10 nm, refractive index ca. 2) in aqueous solution, (ii) the longitudinally coupled SP mode in short linear chains of nanoparticles (diameter 64 nm) coated with CTAB/DNA (estimated refractive index ca. 1.3) and separated by 1 nm gaps and (iii) calculated longitudinal SP mode in linear fused bead strings (diameter 12 nm, no molecular coating, medium is air) as depicted in Fig. 3. The experimental and calculated longitudinal SP modes in Au nanorods red-shift linearly with the increasing aspect ratio. On the contrary, the longitudinally coupled SP mode in assembled nanoparticle chains is sub-linear for aspect ratio exceeding 3 and rapidly saturates as reported also in references [1,2]. The observation of this mode is made possible by the large local optical index of the organic capping layer that shifts the resonance in the range where the dissipative part of the gold dielectric response is minimal (700-800 nm).[13] The calculated red-shifting upon increasing the aspect ratio of the fused bead strings between 1 and 4 is linear but the rate is intermediate between those observed from nanorods and assembled nanoparticle chains.

This observation suggests that longitudinal SP modes in fused bead strings are delocalized but sensitive to the surface corrugation inherited from the native PNN, and still present after fusion, which is rougher that the surface of Au nanorods.

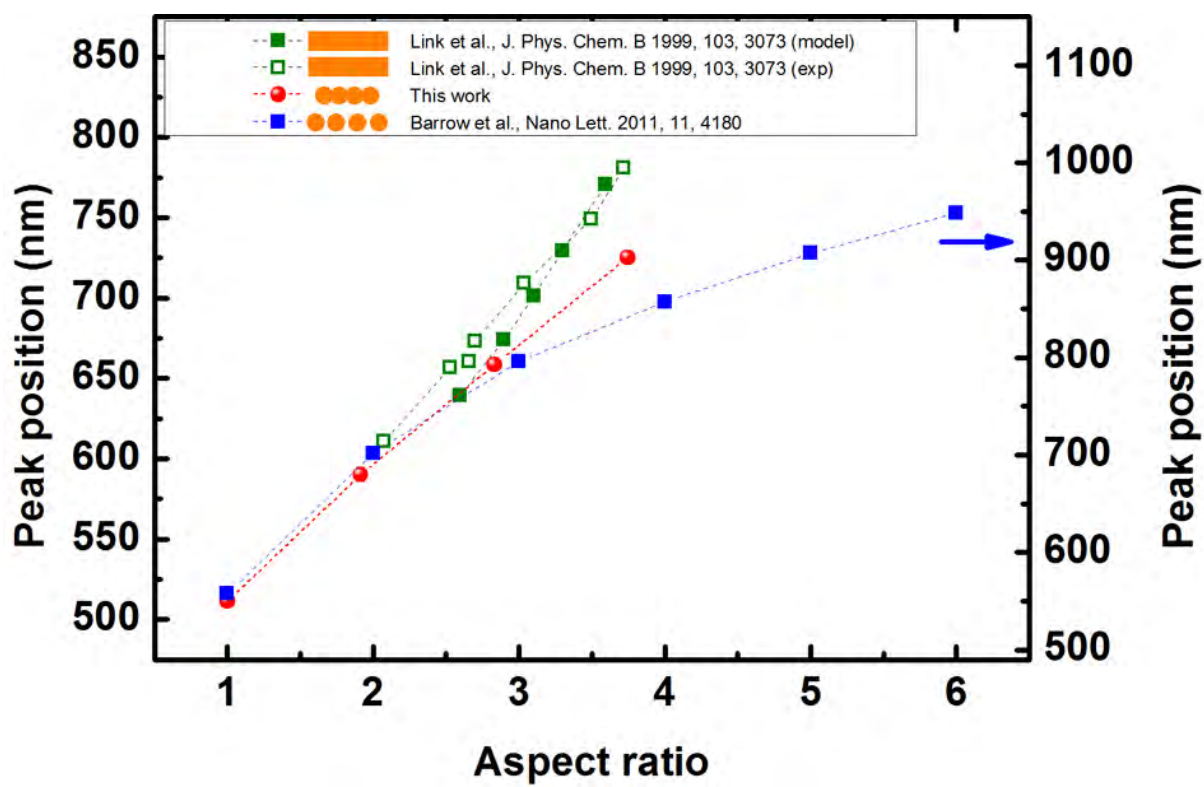

**Figure S5.** Comparison of the aspect ratio evolution of the longitudinal peak resonance in Au nanorods,[14] short and linear nanoparticle chains [15] and fused bead strings.

## 7. Statistical correlation between SP modes and topological features

Self-assembled, and consequently, fused PNN are complex systems, yet they are well-defined both from a synthetic viewpoint and from their optical properties viewpoint at the macroscopic level since the spectroscopic extinction signature of PNN is extremely robust.

The question of associating elementary topological configuration with specific spectroscopic features therefore naturally arises. This issue was considered in our initial work,[1] on the basis of near-field intensity.

Since EELS provides an extremely local spectroscopic probe,[16,17] we could perform a more thorough analysis of spectra associated with basic topological patterns found in all PNN. In particular, PNN can be seen as linear chain fragments cross-linked by Y-shaped junctions and comprising a significant amount of free end chains.

Figure S6 compiles spectra associated with such topological fragments.

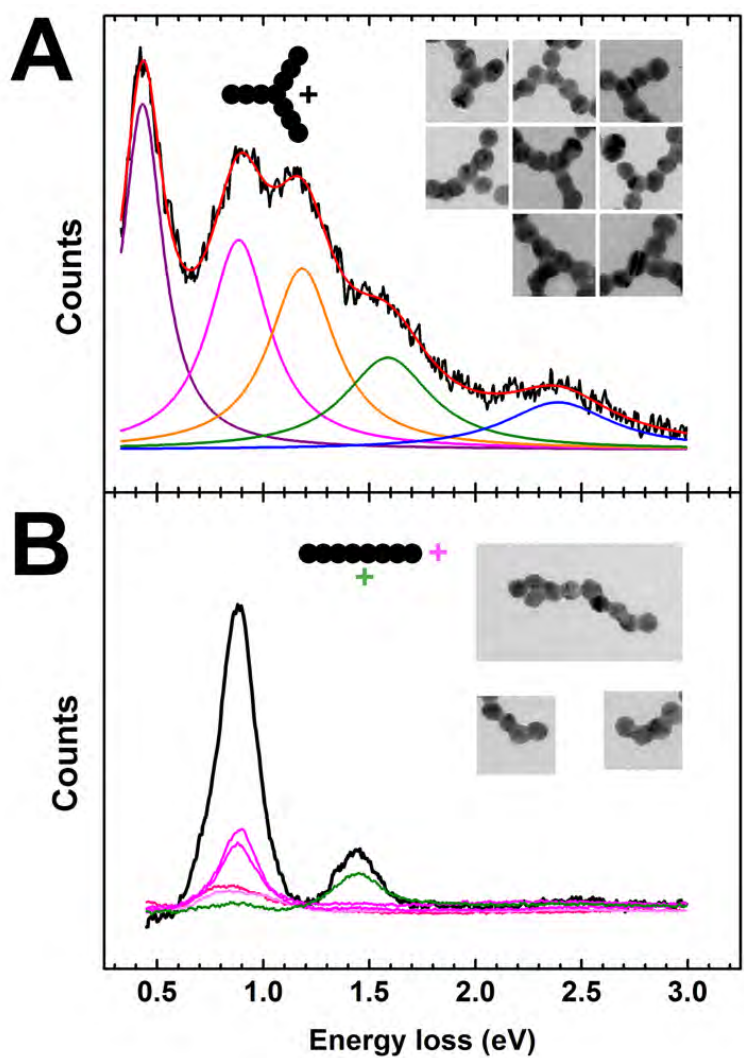

**Figure S6:** (A) Local EELS spectrum obtained by cumulating data specifically from Y-shaped topologies (black line). Insets: TEM images of the Y-shape fragments considered here and schematic representation of the relative position of the probe location (cross) with respect to the structures. In purple, pink, orange, green and blue lines, the five lorentzian peaks optimized so that the sum signal (red line) fits the experimental data. (B) Global (black) and local (pink and green) EELS spectra obtained near linear chains sides (green) and ends (pink). Insets: TEM images of linear chains and free-end chains considered here and schematic representation of the relative probe position.

Spectra associated with Y-shaped junctions were systematically taken from inter-arm regions, about 10 nm from the central nanoparticle as shown by the cross in the schematic inset of Fig. S6A. These spectra exhibit multiple resonances between 2.5 and 0.5 eV, with an increasing intensity for lower energy peaks. In order to attenuate the effect of inhomogeneity in the inter-arm angles and arm lengths, spectra of Y-junctions were accumulated and are shown in Fig. S6 for a set of 8 junctions (black line). This spectrum shows five resonance features, which were associated to lorentzian peak components in agreement with previous EELS spectra analysis.[18] The five lorentzian peaks that give the best fit (red line) are located at 0.43, 0.88, 1.18, 1.59 and 2.39 eV with respective quality factors of 2.10, 2.71, 3.31, 3.29 and 3.94.

The low quality factors are consistent with the residual roughness of the fused nanoparticle chain morphology. It also indicates that merging fused linear chains into a Y junction might favour the emergence of a very low resonance below 0.5 eV, which was not observed in smooth nanorods or in small dendritic nanoparticles.[18]

Indeed, Figure S6B shows spectra associated with linear chains. Experimental frequency-domain spectra of the whole chain exhibit three main peaks at 0.9, 1.5 and 2.5 eV, with intensity increasing rapidly towards lower energy. This is in qualitative agreement with similar spectra of Au nanorods. No peak is found below 0.5 eV. Interestingly, local spectra taken near the central edge of the chain (green cross in schematic inset and green spectra) are predominantly composed of a single peak at 1.5 eV, while local spectra taken near the chain free end (pink cross in inset and pink spectra) contribute mainly to the 0.9 eV feature. This is similar to the excitation of second- and first harmonic plasmon modes, respectively in nanorods (See Fig. 1 in reference [18])

This statistical analysis suggest that the complex LDOS spectra probed by EELS do result from the contribution of specific basic topological units (Y-shaped junctions, free ends and linear fragments) that can often be associated to specific resonances (0.4 eV, 0.9 eV and 1.5 eV respectively).

However, the extreme sensitivity of the LDOS to the specific local arrangement of fused nanoparticles does not permit to build up a simple tinker toy toolbox with which a global spectral signature could be decomposed univocally into a set of elementary topological features.

In fused PNN, the entire complex object should be taken into consideration.

Therefore, we provide more maps of extra PNN that were investigated in Figure S7.
In these maps, the global fusion of the network results in the emergence of long-range delocalized modes that are responsible for the highly contrasted and spectrally tunable distribution of LDOS, which reflects not only in the EELS image but ultimately also in the transmittance of the networks as illustrated in Section 4 for a simple linear chain.

In particular, the three PNNs shown in Panels $A_1$-$A_3$ show bright and localized EELS intensity spots that are markedly displaced upon switching the excitation energy from ca. 0.6 eV (Panels $B_1$-$B_3$) to ca. 1.5 eV (Panels $C_1$-$C_3$). For the lower energies, the spatial distribution of EELS intensity is specific to each structure and varies markedly with the network topology.

Interestingly, we observe a mode convergence at higher energy. Indeed, the higher the selected energy, the more the EELS intensity distribution becomes universal and the less depend on the topology or fusion. This is well illustrated by the maps of the 2.45 eV transverse mode. While the intensity pattern follows closely the network outline, the strong confinement is uniformly distributed near the particles along the entire network for all structures (Panels $D_1$-$D_3$). Panels $E_1$-$E_3$ illustrate once more that our GDM calculations with the discrete dipole approximation account precisely for all the details observed in the experimental maps of this higher energy mode.

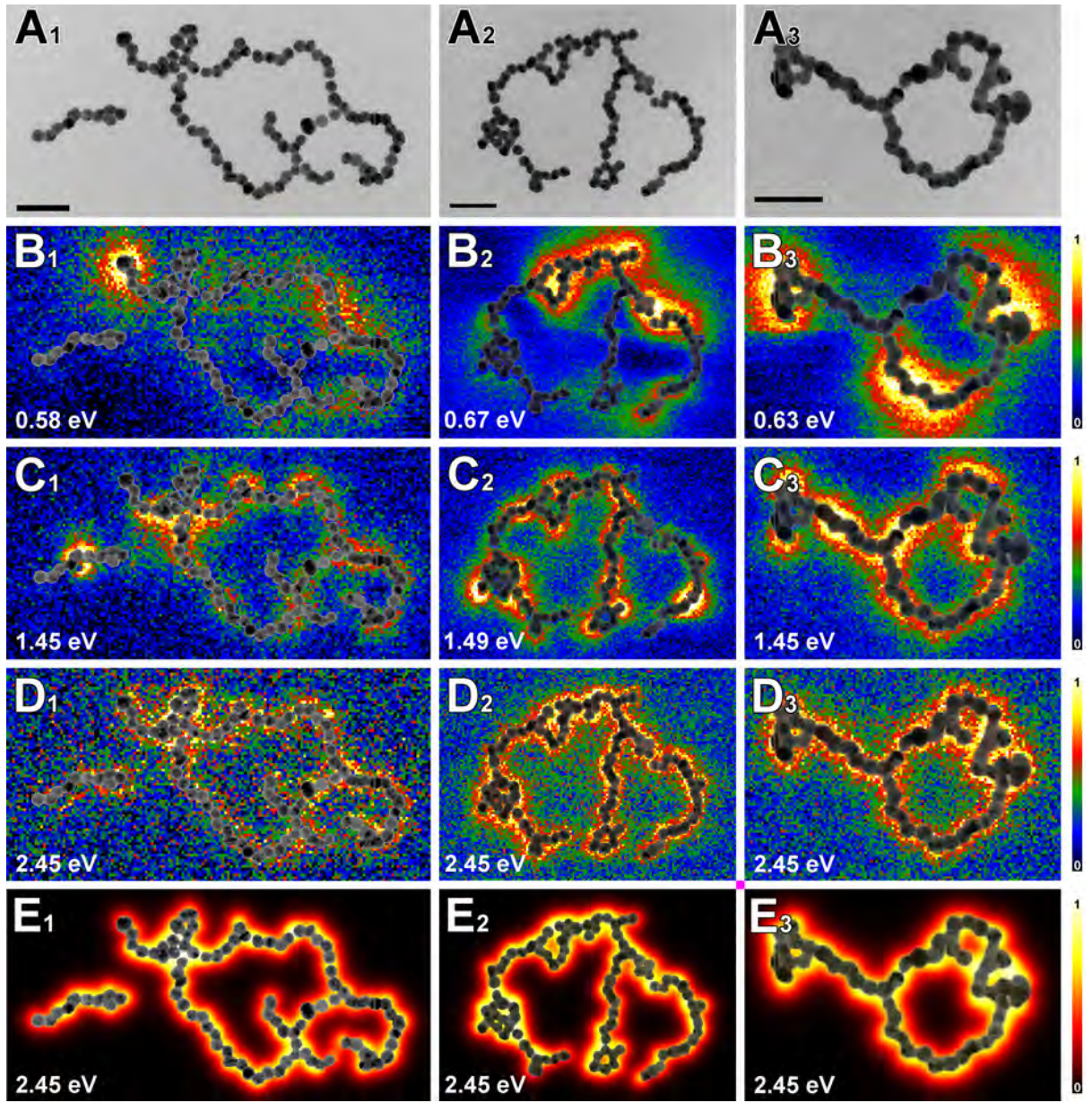

**Figure S7**: Mapping of plasmon modes in complex PNN. (A$_1$-A$_3$) TEM images of fused PNN with complex morphology The small linear fragment aside in A$_1$ was used in the analysis of Fig. S6B. Scale bars are 50 nm. (B-D) Corresponding EELS maps recorded from the well-separated resonance features close to 0.6 eV (B1-B3), 1.45 eV (C$_1$-C$_3$) and at 2.45 eV (D$_1$-D$_3$), with overlaid TEM images of the fused PNN. The same color scale is used for all EELS data and shown on the right side. Data were processed as detailed in Experimental Section. (E$_1$-E$_3$) Simulated EELS map at 2.45 eV computed in the DDA approximation for each of the three structures, the TEM image of which is overlaid. The color scale is shown on the right side.

A. Teulle et al. 2014

## 8. Principal Component Analysis

Principal component analysis is a technique to identify independent spectral features, as linearly uncorrelated "principal components" (PCs).

The first component describes the average EELS intensity in a map (a so-called Score Image), and the average of all spectra from the whole EELS spectrum image (called the first Loading spectrum). The principal components are ordered by the variance that they explain, so the subsequent second, third, etc. components (each consisting of a Score Image and a Loading Spectrum) account for decreasing amounts of spectral variance. After a certain number of extracted principal components, all significant spectral variance in the whole data set is extracted and the remaining principal components just contain random noise. More information and examples can be found in Reference [19].

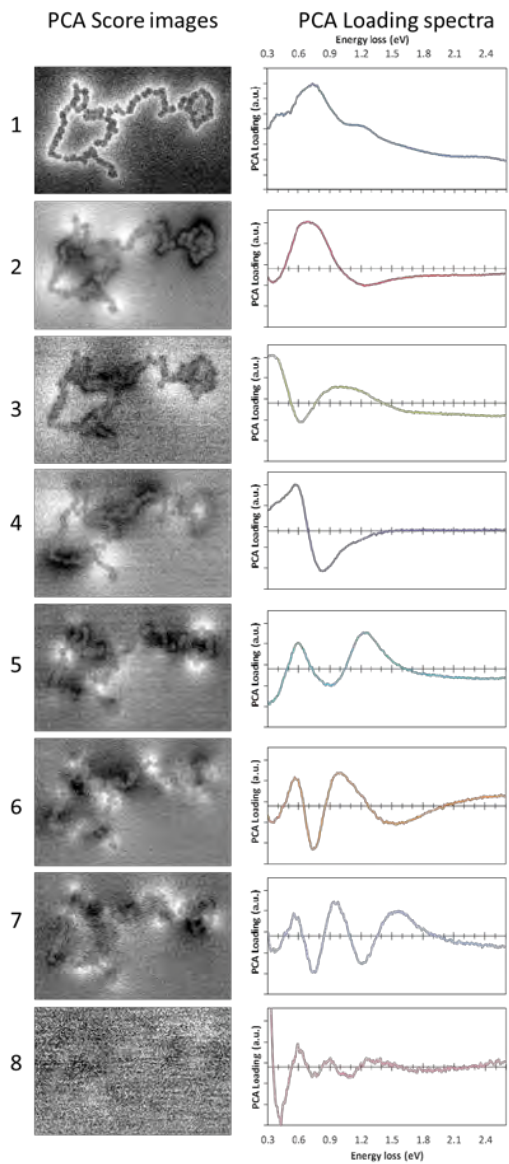

**Figure S8:** Principal Component Analysis of the fused PNN presented in Figure 3

In Figure S9, the PCA results are presented from the spectrum image of the fused PNN presented in Figure 3. The Score Image and Loading Spectrum of the first eight PCs are shown.

From the first PC, it can be seen that the main plasmon peaks occur at 0.4 eV, 0.6-0.8 eV, 1.1 eV and 2.4 eV but the second PC indicates that the 0.6-0.8 eV peaks are stronger but spatially anti-correlated to the other peaks. Note that the Loading Spectrum of the second PC shows which part of the energy range is higher or lower compared to the average spectrum. This confirms our interpretation of Figure S6 in which Y-shaped junctions, linear chains and chain ends contributed differently to the global spectrum, with a linear section and free-end chains indeed contributing more to this component.

The third PC extracts plasmon peaks around 0.4 eV and 1.0 eV (white in the Score Image). Thus, the third and following PCs extract a mix of several plasmon modes, the energies of which are partially overlapping. The eighth and higher PCs mainly extract the random noise in the data set.

This analysis confirms that the number of plasmon modes in our data sets is high with a distinct spatial distribution for the different groups of resonances, which are related to specific local network topologies, in full agreement with our direct analysis of the EELS maps (Figs. 1, 3, 4, S6, S7).

This wealth of both localized and delocalized modes in fused PNN offers a unique opportunity for ultimate waveguide engineering.

A. Teulle et al. 2014

## 9. TEM and EELS map of a 0.7 μm long PNN fragment for energy loss between 0.35 and 2.45 eV

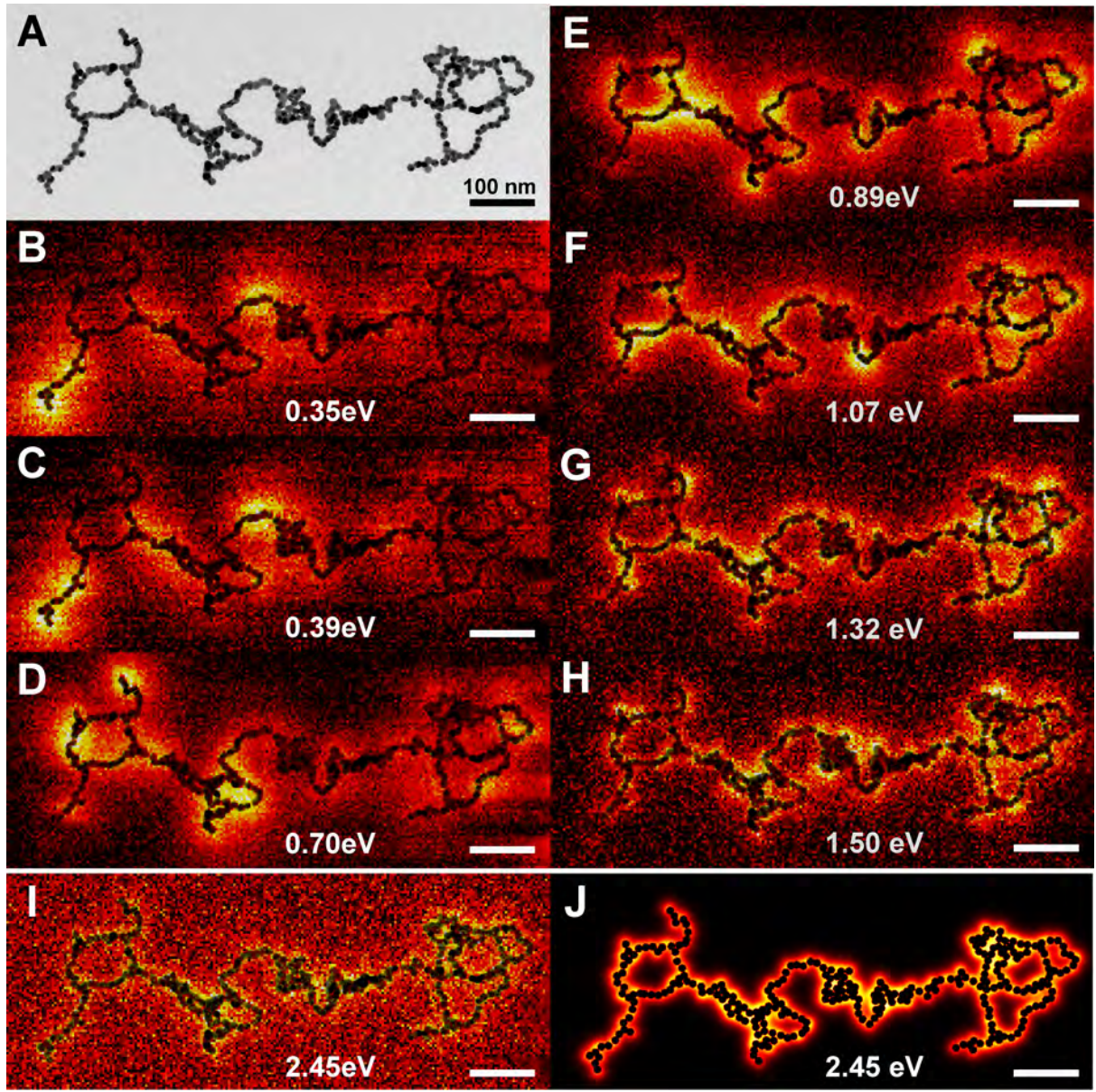

**Figure S9. Plasmon delocalization in extended PNN**. (A) TEM and (B-H) EELS map with TEM overlay of a large nanoparticle chain network recorded at (B) 0.35 eV (3540 nm), (C) 0.39 eV (3180 nm), (D) 0.70 eV (1770 nm), (E) 0.89 eV (1400 nm), (F) 1.07 eV (1160 nm), (G) 1.32 eV (940 nm), (H) 1.50 eV (830 nm). (I) 2.45 eV (506 nm). (J) Simulated EELS map at 2.45 eV. Scale bars: 100 nm. Color scale is similar to Fig. 3B.

EELS mapping in extended fused PNN networks such as the one shown in Fig. S9 shows that low energy modes are characterized by a contrasted series of intense spots located across the entire network. The spatial distribution is significantly modulated as the loss energy is tuned (Panels A-H) with a tendency to be more tightly confined along the particles and more evenly distributed along the chains as the energy reaches the transverse mode at 2.45 eV (Panel I).

The transverse movement of the electrons in this mode is, to a large extent, impervious to the fusion of the particles. This case is thus faithfully accounted for by DDA-type calculations in which the individual nanoparticles are assimilated to point-like dipoles coupled to each other as shown in panel J. The match between experimental (Panel I) and simulated (Panel J) data is excellent. Although, our GDM-based simulation tool is, in principle, able to yield images corresponding to Panels A-H, the required discretization level for these extremely narrow chains combined with the large network size prevents this calculation, due to extremely long computation time.

# References

1. Lin, S., Li, M., Dujardin, E., Girard, C. & Mann, S. One-dimensional plasmon coupling by facile self-assembly of gold nanoparticles into branched chain networks. *Advanced Materials* **17**, 2553-2559 (2005).
2. Li, M., Johnson, S., Guo, H. T., Dujardin, E. & Mann, S. A Generalized Mechanism for Ligand-Induced Dipolar Assembly of Plasmonic Gold Nanoparticle Chain Networks. *Advanced Functional Materials* **21**, 851-859 (2011).
3. Bosman, M. & Keast, V. J. Optimizing EELS acquisition. *Ultramicroscopy* **108**, 837-846 (2008).
4. Martin, O. J. F., Girard, C. & Dereux, A. Generalized Field Propagator for Electromagnetic Scattering and Light Confinement. *Physical Review Letters* **74**, 526-529 (1995).
5. Girard, C., Dujardin, E., Li, M. & Mann, S. Theoretical near-field optical properties of branched plasmonic nanoparticle networks. *Physical Review Letters* **97**, 100801 (2006).
6. Bonell, F. et al. Processing and near-field optical properties of self-assembled plasmonic nanoparticle networks. *Journal of Chemical Physics* **130**, 034702 (2009).
7. Girard, C. & Quidant, R. Near-field optical transmittance of metal particle chain waveguides. *Optics Express* **12**, 6141-6146 (2004).
8. de Abajo, F. J. G. & Kociak, M. Probing the photonic local density of states with electron energy loss spectroscopy. *Physical Review Letters* **100**, 106804 (2008).
9. Cohen-Tannoudji, C., Diu, B. & Laloë, F. *Quantum Mechanics* (Wiley, New York, 1977).
10. Girard, C. Near fields in nanostructures. *Reports on Progress in Physics* **68**, 1883-1933 (2005).
11. Girard, C., Dujardin, E., Marty, R., Arbouet, A. & des Francs, G. C. Manipulating and squeezing the photon local density of states with plasmonic nanoparticle networks. *Physical Review B* **81**, 153412 (2010).
12. Girard, C., Dujardin, E., Baffou, G. & Quidant, R. Shaping and manipulation of light fields with bottom-up plasmonic structures. *New Journal of Physics* **10** (2008).
13. Johnson, P. B. & Christy, R. W. Optical Constants of Noble Metals. *Physical Review B* **6**, 4370-4379 (1972).
14. Link, S., Mohamed, M. B. & El-Sayed, M. A. Simulation of the optical absorption spectra of gold nanorods as a function of their aspect ratio and the effect of the medium dielectric constant. *Journal of Physical Chemistry B* **103**, 3073-3077 (1999).
15. Barrow, S. J., Funston, A. M., Gomez, D. E., Davis, T. J. & Mulvaney, P. Surface Plasmon Resonances in Strongly Coupled Gold Nanosphere Chains from Monomer to Hexamer. *Nano Letters* **11**, 4180-4187 (2011).
16. Batson, P. E. Inelastic-scattering of fast electrons in clusters of small spheres. *Surface Science* **156**, 720-734 (1985).
17. Batson, P. E. Paralell detection for high-resolution electron-energy loss studies in the scanning-transmission electron-microscope. *Review of Scientific Instruments* **59**, 1132-1138 (1988).
18. Bosman, M. et al. Surface Plasmon Damping Quantified with an Electron Nanoprobe. *Scientific Reports* **3** (2013).
19. Bosman, M., Watanabe, M., Alexander, D. T. L. & Keast, V. J. Mapping chemical and bonding information using multivariate analysis of electron energy-loss spectrum images. *Ultramicroscopy* **106**, 1024-1032 (2006).